\documentclass[aps,pre,twocolumn,showpacs,superscriptaddress]{revtex4}

\usepackage{graphicx}
\usepackage[english]{babel}
\usepackage[cp1251]{inputenc}

\begin{document}

\title{One-dimensional steady-state structures at relativistic interaction of laser radiation with overdense plasma for finite electron temperature}

\author{A.~V.~Korzhimanov}
\author{A.~V.~Kim}

\affiliation{Institute of Applied Physics RAS, 603950 Nizhny Novgorod, Russia}

\begin{abstract}
One-dimensional steady-state plasma-field structures in overdense plasma are studied assuming that the electron temperature is uniform over plasma bulk and the ions are stationary. It is shown that there may exist solutions for electron distributions with cavitation regions in plasma under the action of ponderomotive force \end{abstract}


\maketitle

The interaction of superstrong laser radiation with plasma has been recently arousing keen interest for several reasons \cite{mourou}. Firstly, with the advance of laser technologies radiation intensities up to $10^{21} - 10^{22}$ W/cm$^{2}$ have been attained \cite{bahk}. Secondly, wide opportunities open up for using new technologies for practical purposes, such as adronotherapy based on laser accelerated ions \cite{bulanov1}, inertial nuclear fusion \cite{tabak, roth}, generation of coherent X-ray and gamma-spectrum radiation \cite{gordienko}, attosecond pulse generation \cite{naumova}, and others. Finally, this field of research is quite complicated and poorly studied yet, hence there are still many problems to be solved, including fundamental ones. 

It is well known that to provide optical radiation at intensities of about 10$^{18}$ W/cm$^2$ the electron motion in the wave field becomes essentially relativistic, which leads to pronounced changes in the dynamics of interaction of such radiation with plasma. In particular, owing to Lorentz mass increase the effective plasma frequency $\omega_p=(4\pi e^2N_e/m_e)^{1/2}$ ($N_e, m_e$ are the electron concentration and mass, respectively) grows and the plasma that was initially nontransparent at this frequency may become transparent \cite{akhiezer}. Besides, the role of the magnetic component of the Lorentz force acting on the electrons from the direction of the electromagnetic wave increases considerably in the relativistic regime because this force is proportional to the $v_e/c$ ratio, where $v_e$ is electron velocity. This force evokes ponderomotive action of radiation on plasma which, at so high intensities, plays a decisive role in plasma dynamics upon the whole. 

As was mentioned above, the interaction of super-high laser radiation with plasma is a very complicated problem and the main instrument for theoretical research in this field is numerical simulation on super-powerful computers. However, interpretation of results of these studies and problem formulation demand physical intuition based on simple analytical calculations. Therefore, of particular significance in terms of understanding the physics of the processes are the results obtained analytically in some model problems. 

In the current work we study one-dimensional steady-state structures arising in overdense plasma normally irradiated by an intense relativistic laser pulse. This research is a logical continuation of the work done in \cite{marburger, tushentsov, korzhimanov, lai, cattani}, in which analogous studies were carried out in the "cold" plasma approximation when the energy of electron thermal motion was negligible compared to the energy of their oscillatory motion in wave field. Within the above approximation the plasma-field structures were discontinuous as the small parameter of the higher derivative was neglected. In the present work we introduce this small parameter characterizing electron temperature, which enables constructing structures analogous to the obtained earlier but sufficiently smooth. 

In our analysis we will use a joint and self-consistent system of Maxwell's equations for e.m. field and relativistic hydrodynamic equations for electron plasma component. Electron motion will be neglected under the assumption that the ion mass is much higher than the electron mass, so that the times of their reactions are negligible compared to the intervals of the time of interaction of interest to us. For description of e.m. field we will introduce vector $\bf A$ and scalar $\varphi$ potentials and will use the Coulomb gauge ${\rm div}{\bf A}=0$. Then, in the approximation of a steady-state monochromatic circularly polarized plane wave the vector potential may be represented in the following dimensionless form 
\begin{equation}
	\frac{e{\bf A}(z,t)}{m_0c^2}=a(z)e^{i\vartheta(z)}\left({\bf x_0}+i{\bf y_0}\right)e^{i\omega t},	\label{vecpot}
\end{equation}
where ${\bf x_0},{\bf y_0}$ are unit vectors of the $x$- and $y$-axes, $z$ is the axis along which the wave is propagating, $\omega$ is wave frequency, and $m_0$ is electron rest mass. As was shown in \cite{korzhimanov}, Maxwell's equations in plasma in this case may be reduced to the following equations \begin{equation}
	\frac{{\rm d}^2 a}{{\rm d}\zeta^2}-\frac{s^2}{a^3}+\left(1-\frac{n_0 n}{\gamma}\right)a=0	\label{Helmholtz}
\end{equation}
\begin{equation}
	\frac{{\rm d}^2\phi}{{\rm d}\zeta^2}=n_0(n-Z_in_i), \label{Poisson}
\end{equation}
where $\zeta=\omega z/c$; $\phi=e\varphi/m_0c^2$; $s$ is the normalized e.m. energy flux density that is the parameter of the problem; $n_0=4\pi e^2 N_0/m_0 \omega^2$ is the plasma overdense parameter equal to the ratio of a certain characteristic initial electron concentration $N_0$ to its critical value at a given radiation frequency in the linear approximation (hereinafter we will consider only homogeneous plasma layers in which $N_0$ is taken to be the value of unperturbed electron concentration); $Z_i$ is the ion charge (all the ions are supposed to have the same charge); $n=N_e/N_0, n_i=N_i/N_0$ are the normalized electron and ion concentrations, respectively; and $\gamma=\sqrt{1+a^2}$ is electron relativistic factor (i.e., the ratio of their energy to rest energy; thermal energy is neglected here). Thus, the above simplification enabled us to reduce Maxwell's equations to a nonlinear Helmholtz equation for vector potential complex amplitude modulus (\ref{Helmholtz}) and Poisson's equation for scalar potential (\ref{Poisson}). The first of them describes the transverse field produced by the radiation incident on plasma, and the second one describes the longitudinal electrostatic field due to charge separation in plasma. 

The relativistic hydrodynamic equations for a plasma with uniform bulk electron temperature $T_e$ and static ions can be written as
\begin{equation}\label{continuity}
		\frac{\partial N_e}{\partial t}+{\rm div}(N_e {\bf v})=0.
\end{equation}
\begin{equation}\label{hydro}
    \frac{\partial{\bf p}}{\partial t}+({\bf v}\nabla){\bf p}=-T_e\nabla N_e+\frac{e}{c}\frac{\partial{\bf A}}{\partial t}
    +e\nabla\varphi-\frac{e}{c}[{\bf v}\times {\rm rot}{\bf A}]
\end{equation}
\begin{equation}\label{momentum}
		{\bf p}=\gamma m_0 {\bf v}.
\end{equation}
Here ${\bf v},{\bf p}$ are the mean velocity and pulse of the electron component, respectively. This system may also be reduced in one-dimensional geometry. As a result, in the stationary case it reduces to two equations, the first of which is the law of conservation of a canonic electron pulse in transverse direction:
\begin{equation}\label{canon_momentum}
    {\bf p}_\bot-\frac{e}{c}{\bf A}=0.
\end{equation}
The second equation is a simple balance of forces acting on the electrons:
\begin{equation}\label{force_balance}
    \frac{{\rm d}\phi}{{\rm d}\zeta}-\frac{{\rm d}\gamma}{{\rm d}\zeta}-\mu\frac{1}{n}\frac{{\rm d}n}{{\rm d}\zeta}=0.
\end{equation}
The first term here corresponds to the electrostatic force of charge separation in plasma, the second to the ponderomotive force of the laser field, and the third term corresponds to the internal pressure forces arising due to temperature ($\mu=T_e/m_0c^2$ is the parameter characterizing electron temperature). The continuity equation (\ref{continuity}) is fulfilled automatically in this case, thus ensuring plasma quasineutrality.

\begin{figure}[htb]
	\includegraphics[width=.5\textwidth]{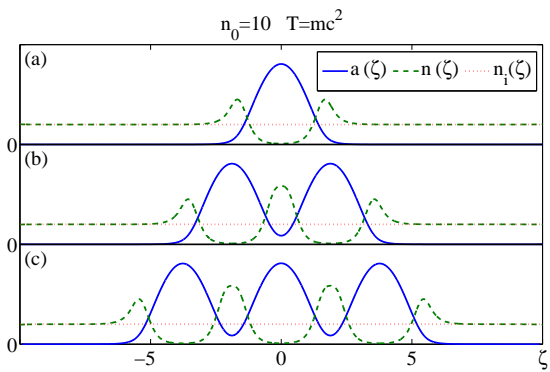}
	\caption{Examples of localized solutions of system (\ref{Helmholtz_inf}) - (\ref{Poisson_inf}). (a) First mode, one field antinode. (b) Second mode, two field antinodes. (c) Third mode, three field antinodes.} \label{solution_infinite}
\end{figure}

We first consider solutions to the system of equations (\ref{Helmholtz}),(\ref{Poisson}) and (\ref{force_balance}) in infinite homogeneous plasma ($Z_in_i=1$ throughout the space). In this case, it is convenient to integrate equation (\ref{force_balance}) and express the concentration through
\begin{equation}\label{conc_infinite}
    n=\exp{\frac{\phi-\gamma+1}{\mu}},
\end{equation}
where it is implied that we are interested only in localized solutions for which $a\rightarrow 0, \phi\rightarrow 0, n\rightarrow 1$ is fulfilled for $\zeta\rightarrow\pm\infty$. In addition, $s=0$ also holds for localized solutions in infinite plasma. Substituting (\ref{conc_infinite}) into the two remaining equations yields the following fourth-order system of ordinary differential equations 
\begin{eqnarray}
	\frac{{\rm d}^2 a}{{\rm d}\zeta^2}&=&\left(\frac{n_0}{\sqrt{1+a^2}}
	\exp{\frac{\phi-\sqrt{1+a^2}+1}{\mu}}-1\right)a	\label{Helmholtz_inf}\\
	\frac{{\rm d}^2\phi}{{\rm d}\zeta^2}&=&n_0\left(\exp{\frac{\phi-\sqrt{1+a^2}+1}{\mu}}-1\right), \label{Poisson_inf}
\end{eqnarray}
Unfortunately, the above system has no analytical solution, but we found one of its integrals that is written as 
\begin{eqnarray}
	\frac{1}{2}\left(\frac{{\rm d}a}{{\rm d}\zeta}\right)^2+\frac{1}{2}a^2-
			\frac{1}{2}\left(\frac{{\rm d}\phi}{{\rm d}\zeta}\right)^2-n_0\phi-\nonumber\\
			-\mu n_0\exp{\frac{\phi-\sqrt{1+a^2}+1}{\mu}}=const. \label{integral}
\end{eqnarray}
This integral simplifies search for localized solutions to our problem. Indeed, let us set a problem of finding localized solutions symmetric relative to the $\zeta=0$ plane. The relation 
$$
	\left.\frac{{\rm d}a}{{\rm d}\zeta}\right|_{\zeta=0}=0\ \  \left.\frac{{\rm d}\phi}{{\rm d}\zeta}\right|_{\zeta=0}=0
$$
is fulfilled for such solutions. And the relationship between the values of vector and scalar potentials at the same point gives the integral (\ref{integral}):
\begin{eqnarray}
	\frac{1}{2}a^2(0)-n_0\phi(0)-\nonumber\\
	\mu n_0\left(\exp{\frac{\phi(0)-\sqrt{1+a^2(0)}+1}{\mu}}-1\right)=0.\nonumber
\end{eqnarray}
We now have one fitting parameter (for example, $\phi(0)$), by changing which and solving the system (\ref{Helmholtz_inf}) - (\ref{Poisson_inf}) we find a localized solution. Our studies showed that in such a formulation there may exist a family of solutions differing by the number of field antinodes. Examples for the first and third modes are given in Fig.\ref{solution_infinite}. Note that, for inexplicable reason, we failed to construct modes higher than the first one for small values of parameter $\mu$. This is most probably a result of numerical errors made in constructing the solution, but it may also have physical explanation. 

Besides, the integral (\ref{integral}) allows us to show that there are no antisymmetric solutions within the framework of our system for which the relation 
$$
	\left.a\frac{}{}\right|_{\zeta=0}=0\ \  \left.\frac{{\rm d}\phi}{{\rm d}\zeta}\right|_{\zeta=0}=0.
$$
is fulfilled in the $\zeta=0$ plane. At the same time, we constructed such solutions in a cold plasma (see, e.g., \cite{korzhimanov}). This isn't surprising either physically or mathematically. Indeed, in terms of mathematics we increased the order of the studied system, which may lead to appreciable changes in solution topology and phase plane upon the whole. In terms of physics, introduction of nonzero temperature results in appearance of electrons that can move from one field antinode to another during heat motion, thus providing their interaction. This interaction most likely leads to the nonstationarity of antisymmetric distributions. 

Solutions constructed in an infinite plasma point to some principal features of the regime of interaction of a superintense laser pulse with overdense plasma. These are appreciable redistribution of electron concentration under the action of ponderomotive force and the related longitudinal electrostatic field approximately equal to the radiation electric field, as well as existence of structures with several electron layers separated by gaps almost devoid of electrons (fig.~\ref{solution_infinite}b).

Let us now pass over to the case of a bounded plasma layer that is more important for practical reasons. We will restrict our consideration to the one-sided irradiation of plasma. Then, it is convenient to construct solutions starting at some point outside plasma, from the unirradiated side rather far from the plasma so that the conditions $n, \phi^{\prime}, a^\prime << 1$ (the primes mean differentiation along the coordinate) can be regarded to be met at this point. Note that $a\approx\sqrt{s}$, and the function $\phi(\zeta)$ is generally unbounded (in the one-dimensional geometry the potential is known to behave logarithmically at large distances). This makes the system (\ref{Helmholtz_inf}) - (\ref{Poisson_inf}) unhandy for constructing solutions. We remind the reader that this system was obtained from the basic equations (\ref{Helmholtz}), (\ref{Poisson}) and (\ref{force_balance}) by omitting the variable $n$. By introducing a new variable $e$ that denotes a longitudinal electrostatic field and is define by 
$$
	e=-\frac{{\rm d}\phi}{{\rm d}\zeta}
$$
we obtain an equivalent system of equations. In this fashion we omit from the initial system of equations variable $\phi$ for which boundary conditions are not clear and obtain the following system 
\begin{eqnarray}
	\frac{{\rm d}^2 a}{{\rm d}\zeta^2}&=&\frac{s^2}{a^3}-\left(1-\frac{n_0n}{\sqrt{1+a^2}}\right)a	\label{Helmholtz_fin}\\
	\frac{{\rm d}e}{{\rm d}\zeta}&=&n_0\left(Z_in_i(\zeta)-n\right) \label{Poisson_fin}\\
	\mu\frac{{\rm d}n}{{\rm d}\zeta}&=&-n\left(e+\frac{{\rm d}\sqrt{1+a^2}}{{\rm d}\zeta}\right) \label{force_bal_fin}
\end{eqnarray}
Here $n_i(\zeta)$ is a given function specifying initial plasma distribution in space. We will further restrict our consideration to the case of a plane homogeneous layer of thickness $L$ so that this function will have the form 
\begin{equation}\label{ions}
	n_i(\zeta)=\left\{
					\begin{array}{cll}
					0 & \quad{\rm for}\quad \zeta<0,\; \zeta>\omega L/c,\\
					1/Z_i & \quad{\rm for}\quad 0<\zeta <\omega L/c.\\
					\end{array}
				\right.
\end{equation}
By analogy with (\ref{Helmholtz_inf}) - (\ref{Poisson_inf}) the system (\ref{Helmholtz_fin}) - (\ref{force_bal_fin}) has an integral analogous to (\ref{integral}) which, in the case under consideration, has the following form 
\begin{eqnarray}
	\frac{1}{2}\left(\frac{{\rm d}a}{{\rm d}\zeta}\right)^2+\frac{1}{2}a^2-\frac{1}{2}\left(\frac{s}{a}\right)^2-
		\frac{1}{2}e^2-\mu n_0 n- \nonumber\\
		-n_0Z_i\int en_i(\zeta){\rm d}\zeta = {\rm const}. \label{integral_2}
\end{eqnarray}
Unfortunately, for $n_i(\zeta)\neq 0$ the integral (\ref{integral_2}) is not an algebraic expression and contains integration in explicit form. This restricts its applicability. However, it may be used in the region fully devoid of ions, where the last term in the left-hand side of the expression (\ref{integral_2}) is identically equal to zero.

As was mentioned above, we will construct solutions starting at some point $\zeta_{st}$ that is rather far from the plasma layer from its unirradiated side. Suppose that $a=\sqrt{s}+\alpha$ at this point and the inequalities $\alpha(\zeta_{st})<<1$, $n(\zeta_{st})<<1$, $e(\zeta_{st})<<1$ are fulfilled. In addition, $n_i(\zeta_{st})=0$ as this point is outside the plasma layer. Then, in the neighborhood of the point $\zeta_{st}$ we can linearize equations (\ref{Helmholtz_fin}) - (\ref{force_bal_fin}) in a standard manner to obtain \begin{eqnarray}
	\frac{{\rm d}^2 \alpha}{{\rm d}\zeta^2}&=&-4\alpha+n_0 n\sqrt{\frac{s}{1+s}} \label{linear_1}\\
	\frac{{\rm d}e}{{\rm d}\zeta}&=&-n_0 n \label{linear_2}\\
	\mu\frac{{\rm d}n}{{\rm d}\zeta}&=&-n\left(e+\sqrt{\frac{s}{1+s}}\frac{{\rm d}\alpha}{{\rm d}\zeta}\right) \label{linear_3}
\end{eqnarray}
Note that it would be wrong to neglect in the third equation quantities of the form $ne$ and $n{\rm d}\alpha/{\rm d}\zeta$ that are small compared to ${\rm d}n/{\rm d}\zeta$ because the resulting system would have only a trivial solution; hence we preserve these terms. As the system (\ref{linear_1}) - (\ref{linear_3}) is still nonlinear, its solution is rather complicated, but assuming that the inequality 
\begin{equation}\label{ineq}
	e>>\sqrt{\frac{s}{1+s}}\frac{{\rm d}\alpha}{{\rm d}\zeta}
\end{equation}
is met, equations (\ref{linear_2}) and (\ref{linear_3}) are separated from (\ref{linear_1}), which allows finding their solutions. Indeed, if we divide the second of these equations by $n$, differentiate it and substitute ${\rm d}e/{\rm d}\zeta$ from the first equation, we will obtain one second-order equation for $n$:
\begin{equation}\label{conc_diff_eq}
	\mu \frac{{\rm d}^2 n}{{\rm d}\zeta^2}-\mu\frac{1}{n}\left(\frac{{\rm d} n}{{\rm d}\zeta}\right)^2-n_0n^2=0.
\end{equation}
Let us seek solution to this equation in the form
$$
	n=C(\zeta-\zeta_0)^\beta
$$
and obtain
$$
	\begin{array}{rl}
		\mu C\beta\left(\beta-1\right)(\zeta-\zeta_0)^{\beta-2}-\mu C \beta^2 (\zeta-\zeta_0)^{\beta-2}-&\\
		-n_0 C^2(\zeta-\zeta_0)^{2\beta}&=0.
	\end{array}
$$
This equality must be identically fulfilled for all values of $\zeta$; therefore, the conditions 
$$
	\beta=-2,	\qquad	C=\frac{2\mu}{n_0}
$$
must be met. Finally the solution of equation (\ref{conc_diff_eq}) takes on the form
\begin{equation}\label{n_solution}
	n(\zeta)=\frac{2\mu}{n_0(\zeta-\zeta_0)^2}.
\end{equation}
From equation (\ref{linear_2}) we find
\begin{equation}\label{e_solution}
	e(\zeta)=\frac{2\mu}{\zeta-\zeta_0},
\end{equation}
And from (\ref{linear_1}) follows
\begin{equation}\label{a_solution}
	\alpha(\zeta)=\frac{\mu}{2(\zeta-\zeta_0)}\sqrt{\frac{s}{1+s}}.
\end{equation}
We can now readily verify that the inequality (\ref{ineq}) that was supposed to be met does hold sufficiently far from the point $\zeta=\zeta_0$. Note that all the integration constants in the derivation of (\ref{n_solution}) - (\ref{a_solution}) were chosen so that the quantities $\alpha$, $e$, $n$ vanished at the infinity.

Generally speaking, in the construction of the solution $\zeta_0$ is an unknown quantity that determines a whole family of solutions. There arises a question which of these solutions is correct. The answer to this question is quite trivial. One must choose the solution at which the conditions of plasma quasineutrality upon the whole are fulfilled:
\begin{equation}\label{quasineutrality}
	\int\limits^{+\infty}_{-\infty}\left(Z_in_i(\zeta)-n(\zeta)\right){\rm d}\zeta=0.
\end{equation}
Actually, the problem of constructing a solution reduces to sorting all solutions by parameter $\zeta_0$ and finding the one that would satisfy the condition (\ref{quasineutrality}). Moreover, there is no need to introduce $\zeta_0$ in this case. Indeed, from (\ref{n_solution}) - (\ref{a_solution}) we can find the following relations
\begin{eqnarray}
	n & = & \frac{e^2}{2\mu n_0} \label{n_condition}\\
	\alpha & = & \frac{e^2}{8\mu} \sqrt{\frac{s}{1+s}} \label{a_condition}\\
	\frac{{\rm d}\alpha}{{\rm d}\zeta} & = & -\frac{e^3}{8\mu^2}\sqrt{\frac{s}{1+s}} \label{dadz_condition}
\end{eqnarray}
Thus, we have excluded parameter $\zeta_0$.

Now the solution procedure will be the following. We choose a certain point $\zeta_{st}>\omega L/c$ at which we set the value of variable $e$. The values of $a$, ${\rm d} a/{\rm d}\zeta$ and $n$ at this point are found from the relations (\ref{n_condition}) - (\ref{dadz_condition}) taking into account that $a=\sqrt{s}+\alpha$. Thus, we have boundary conditions for the system (\ref{Helmholtz_fin}) - (\ref{force_bal_fin}) at the point $\zeta_{st}$. We solve the system of equations and check whether the equality (\ref{quasineutrality}) is fulfilled. If it is fulfilled, then it is the required solution; if the inequality is not fulfilled, we change the value of the variable $e$ at the point $\zeta_{st}$ and repeat the procedure. We proceed in this fashion until we find the right solution. In view of complexity of the system its solution was done numerically by manually sorting the value of $e(\zeta_{st})$. (There are some ideas of how this sorting might be automated, but they have not been implemented for lack of the barest necessity.)

Now a few words about parameter $s$. As we have mentioned above, it stands for the e.m. energy flux density and is equal to the intensity of radiation passing through the layer. In our problem it determines the amplitude of the wave incident on the layer. Thus, if we have to construct a function of the incident wave amplitude, then we must sort parameter $s$ for each of its values using the procedure described above. 

\begin{figure}[htb]
	\includegraphics[width=.5\textwidth]{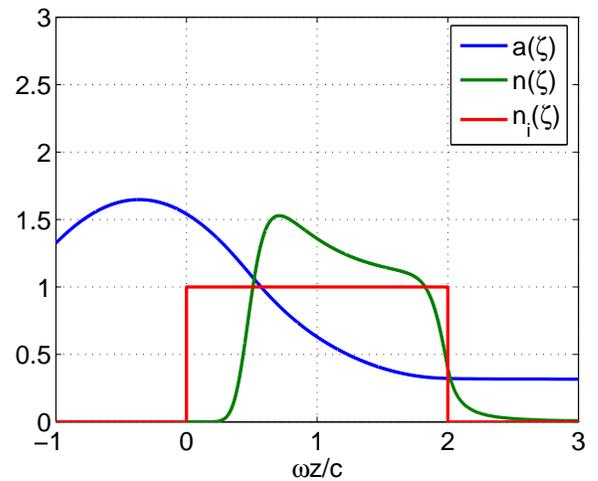}
	\caption{Example of solution of system (\ref{Helmholtz_fin}) - (\ref{force_bal_fin}) in a plasma layer of finite thickness. Parameters of the problem: $\mu=0.01$, $n_0=2$, $\omega L/c=2$, $s=0.1$. The corresponding amplitude of incident radiation $a_{i}=0.85$.}\label{solution_finite}
\end{figure}

An example of solution constructed by the described algorithm is presented in fig.~\ref{solution_finite}. The parameters of the problem were taken to be $n_0=2$, $s=0.1$, $\omega L/c=2$. As soon as the solution has been constructed one can find the amplitude of the incident pulse which, in a real situation, is a true parameter of the problem specified in the course of experiment. For this the values of $a$ and ${\rm d}a/{\rm d}\zeta$ must be known at an arbitrary point in vacuum from the irradiated side of the plasma layer. Indeed, at this point there are incident and reflected waves, hence we can write the following equality 
\begin{equation}\label{inc_and_refl}
	a(\zeta)e^{i\vartheta(\zeta)}=\hat{a}_{i}e^{-i\zeta}+\hat{a}_{r}e^{i\zeta},
\end{equation}
where $\vartheta(\zeta)$ was introduced in equation (\ref{vecpot}) and $\hat{a}_{i}$ and $\hat{a}_{r}$ denote complex amplitudes of the incident and reflected waves, respectively. By differentiating equation (\ref{inc_and_refl}) and omitting from the resulting system the quantity $\hat{a}_{r}$ we obtain 
\begin{equation}\label{inc_ampl_eq}
	\left[\frac{{\rm d}a}{{\rm d}\zeta}+ia\left(\frac{{\rm d}\vartheta}{{\rm d}\zeta}-1\right)\right]e^{i\vartheta}=
		-2\hat{a}_{i}e^{-i\zeta}.
\end{equation}
Taking the modulus of this expression and bearing in mind that parameter $s$ was introduced in the form (see \cite{korzhimanov})
$$
	s=-a^2\frac{{\rm d}\vartheta}{{\rm d}\zeta}
$$
we finally obtain an expression for the real amplitude of the wave incident on the layer: 
\begin{equation}\label{inc_ampl}
	a_i=\frac{1}{2}\left[\left(\frac{{\rm d}a}{{\rm d}\zeta}\right)^2+a^2\left(\frac{s}{a^2}+1\right)^2\right]^{1/2}.
\end{equation}

\begin{figure}[htb]
	\includegraphics[height=6 cm,width=8 cm]{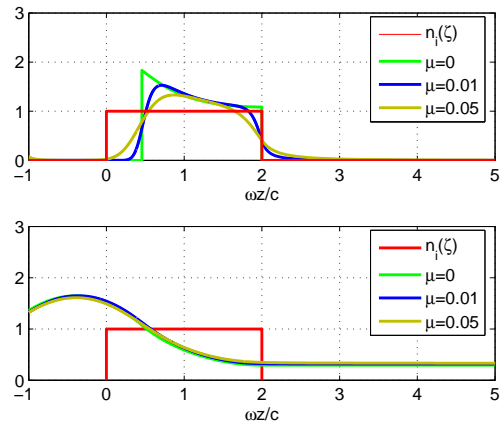}
	\caption{Comparison of solutions of system (\ref{Helmholtz_fin}) - (\ref{force_bal_fin}) for different values of $\mu$ but equal incident wave amplitudes. Distributions of electron concentration are given on top, and of laser field at the bottom. The solution for $\mu=0$ was constructed by the method described in \cite{korzhimanov}.} \label{comprehension_different_mu}
\end{figure}

It is also interesting to compare solutions with different values of electron temperature $\mu$. If parameter $s$ is fixed, then the value of the incident wave amplitude will be different for different values of $\mu$. Therefore, such a comparison will not be physically correct. For a more correct comparison we fitted parameter $s$ to each value of $\mu$ so as to ensure equal incident wave amplitudes in all the cases. The result is demonstrated in fig.~\ref{comprehension_different_mu}. Note that the solutions obtained for $\mu=0$ by the method described in \cite{korzhimanov} are also shown in this figure. The characteristic feature of this solution is enhancement of electron layer border blurring with increasing temperature, which is a natural and expected effect. Upon the whole, it should be noted that solutions of the system (\ref{Helmholtz_fin}) - (\ref{force_bal_fin}) with $0<\mu<0.01$ differ from the solutions with $\mu=0$ obtained in the work \cite{korzhimanov} only slightly and are comparable in terms of complexity of their construction. This enables us to conclude that both the approaches to finding steady-state distributions are equivalent. The only drawback of the solutions with nonzero temperature is divergence of the value of scalar potential at the infinity. However, this divergence is logarithmic and may be eliminated manually by truncating solutions at some distance from the plasma layer. 

\begin{figure}[htb]
	\includegraphics[height=6 cm,width=8 cm]{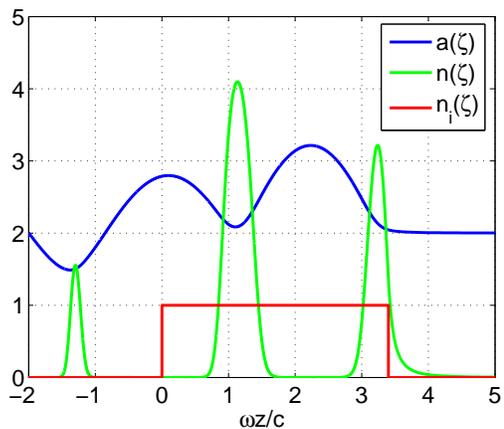}
	\caption{Example of solution of system (\ref{Helmholtz_fin}) - (\ref{force_bal_fin}) with three electron layers. Layer parameters: $\mu=0.01$, $n_0=2$, $\omega L/c=3.4$; incident wave amplitude $a_i=2.1$.} \label{two_layers}
\end{figure}

In \cite{korzhimanov} it was shown that the so-called resonator-like structures in which electron layers separated merely by an ion gap play the role of walls may exist in a layer of finite thickness. Generally speaking, structures with an arbitrary number of electron layers separated by ion gaps may exist in a plasma layer. However, a special algorithm is needed for constructing solutions with $N$ electron layers as such solutions are discontinuous, with the number of solution discontinuities being proportional to $N$. This circumstance strongly complicates construction of such solutions, whereas under the action of temperature the solutions become smooth and all the family may be obtained identically by simply varying the values of $s$ and $e(\zeta_{st})$. An example of solution with three electron layers one of which is outside the plasma layer is given in fig.~\ref{two_layers}. 

To conclude, we proposed a method for constructing steady-state structures in a plane plasma layer irradiated by superintense laser pulse. Introduction of uniform bulk electron temperature enables us to obtain smooth solutions. In spite of the fact that the problem is a model one, the obtained solutions demonstrate basic features of overdense plasma dynamics in the field of superintense waves, such as electron density redistribution under the action of ponderomotive forces, the presence of appreciable electrostatic field in the longitudinal direction, and possible existence of resonator-like solutions with several electron layers separated by relatively rarefied gaps. The presented solutions may also be used for analysis of possible ways of producing strong longitudinal fields in solid-state targets, as was discussed in, e.g., \cite{macchi,kor2}. Besides, these solutions may be a handy tool for analysis of the phenomenon of relativistically induced transparency \cite{cattani}.

\end{document}